\documentclass[a4paper]{article}

\usepackage{amssymb}
\usepackage{amsmath}
\usepackage{latexsym}
 
\def\uc  {universal covering}

\def\setC{\mathbb{C}}

\def\ie {{i.e.}}\def\calT {{\cal T}}

\def\R{{\rm I\!R}}
\def\bbbN{{\rm I\!N}}

\def\npage {\vfill \eject}
\def\lr{Lachi\`eze-Rey}

\newcommand{\matrixdd}[4]{\left[\begin{array}{cc}#1&#2\\#3&#4\\ \end{array} \right]}

\title{Eigenmodes  of Dodecahedral space}
\author{M. Lachi{\`e}ze-Rey\\
Service d'Astrophysique, C. E. Saclay\\91191 Gif sur Yvette cedex, 
France}

\begin{document}
\maketitle
\abstract{Cosmological models where spatial sections are the Poincar\'e dodecahedral space D have been recently invoked to give an account of the lower modes of the angular anisotropies of the cosmic microwave background.  Further explorations of  this possibility require the knowledge of the eigenmodes of the Laplacian of D. Only the first modes have been calculated numerically.
Here we give an explicit form for these modes up to arbitrary order, in term of eigenvectors of a small rank matrix, very easy to calculate numerically. As an illustration  we give the first   modes, up to the eigenvalue $-k (k+2)$ for $k=62$.  These results  are obtained by application of a more general method (presented in a previous work)  which allows to express the properties of any eigenfunction of the Laplacian of the  three sphere under an arbitrary rotation of SO(4).
}
\section{Introduction}
There is a long time interest for cosmological models where space is multi- rather than simply-connected (see a review in \cite{Lalu}). Recently,   \cite{Luminet} claimed that a peculiar model may give an account of the two first moments in the angular power spectrum of the anisotropies of the cosmic microwave background. This model involves the Poincar\'e dodecahedral space $\mathbf{ D}$, whose \uc ~is the three sphere $S^3$.
The  calculations  involved    the first  eigenmodes [of the Laplacian] in $\mathbf{ D}$, which were estimated by numerical methods. However, to make predictions beyond the first moments, and to check non diagonal terms in the correlation matrix, the knowledge of a greater number of modes is required. 

The  eigenmodes of a spherical space $S^3/\Gamma$ are the  an eigenmodes of  $S^3$ which are conserved by all the rotations of $\Gamma$. 
In general, they  remain unknown. Their number, \ie, the multiplicity of the eigenspaces, 
was calculated   by \cite{Ikeda}  as a function of the eigenvalue of the Laplacian. 
Recently,   \cite{Leh} (see also \cite{Leh2}) have provided analytic  calculations of these modes for Lens space and Prism space. The same  results were obtained later with  a different  method by   \cite{lachs3}.
Here we provide a new and efficient method to calculate  the  modes of the dodecahedral space. 
In  \cite{lachs3}, we constructed a special basis (B3) for the modes of $S^3$, which allowed to calculate explicitely their behaviour under any rotation of SO(4). These results are recalled in section 2. They are then applied to $\mathbf{ D}$ (section 3), which leads to the derivation of its  eigenmodes.
 
\section{Rotation properties of the eigenmodes of $S^3$}
 
The eigenvalues of the Laplacian $\Delta$ of $S^3$ are of the form $\lambda _{k} = -k~(k+2)$,  where $k \in \bbbN^+$.     For a given value of $k$, they span the  eigenspace 
    ${\cal V} ^{k }$ of   dimension $(k+1)^2$. 
  The eigenmodes of the   dodecahedral space $\mathbf{ D} \equiv S^3/G $ are those eigenmodes of $S^3$ which remain invariant under all holonomy transformations of $G$. To be so, it is necessary and sufficient that they are   invariant under the two generators of $G$, that we call $g_{\pm}$.   
    
We will use a specific basis B2   of ${\cal V} ^{k }$, whose properties were detailed in \cite{lachs3}.
This basis   was originally introduced by \cite{bander}. A real version of  it   has been used   by \cite{Leh} to find the eigenmodes of lens and prism spaces.  It  is generated by a set of  $(k+1)^2$  functions:\begin{equation} \label{B2}
B2 \equiv ( T_{k;m_1,m_2}) ,~ m_1, m_2=-k/2...k/2,
\end{equation} where  $m_1$ and $m_2$ vary  independently    by entire increments. 
We know (\cite{Ikeda}) that $\mathbf{ D}$ has eigenmodes only for $k$ even, that we assume from now. This ensures that $m_1$ and $m_2$ are entire. Hereafter, all sums involving $m_1$ or  $m_2$ will cover the range $-k/2..k/2$.
 
    These functions are naturally adapted to   toroidal   coordinates to describe $S^3$: 
      $(\chi, \theta, \phi)$ spanning  the range $0 \le \chi \le \pi/2$, $0 \le \theta \le 2\pi$ and $0 \le \phi \le 2\pi$. 
 They are conveniently defined 
 (see \cite{Leh} for a more complete description) from an   isometric  embedding of $S^3$  in $\R ^4$ (as the hypersurface $x \in \R ^4; ~\mid x \mid =1$):
 \begin{displaymath}
      \left\{
      \begin{array}{ccc}
          x^0 & = & r~\cos \chi ~\cos \theta \\
          x^1 & = & r~\sin \chi ~\cos \phi   \\
          x^2 & = & r~\sin \chi ~ \sin \phi  \\
 x^3 & = & r~\cos \chi ~\sin \theta \\
     \end{array}
      \right.
\end{displaymath}
 where  $ (x^{\mu}),~\mu=0,1,2,3$  is a point of $\R ^4$. 
 
    In \cite{lachs3}, we gave the explicit expression of B2:   \begin{equation} 
\label{B2modes}
T_{k;m_1,m_2} (X )= C_{k;m_1,m_2}~  [\cos \chi ~ e^{i \theta}] ^{\ell  } ~[\sin \chi ~ e^{i \phi}] ^{m  }  ~P^{( m,\ell)}_d[\cos (2 \chi)]~.\end{equation} We wrote 
$\ell =m_1 + m_2, ~m=m_2 - m_1$ for simplicity;   $P^{( m,\ell)}_d$ is a Jacobi polynomial.
Normalization leads to    $C_{k;m_1,m_2}~\equiv \frac{\sqrt{ (k+1)}}{\pi} ~\sqrt{\frac{ (k/2+m_2)!~(k/2-m_2)! }{(k/2+m_1)!~Ê(k/2-m_1)!}}$.    
    
    Note that the    basis functions $T_{k;m_1,m_2}$  have  also been  introduced in \cite{erd} (p. 253), with their expression in Jacobi Polynomials.  They are the complex counterparts of those proposed by \cite{Leh}  (their equ. 19). The variation range of the indices  $m_1,m_2$ here is equivalent to 
 their condition \begin{equation} \label{ }
\mid \ell \mid + \mid m \mid  \le k,~\ell +m =k, ~\mod{2},
\end{equation}
through the correspondence $\ell=m_1+m_2,~m=m_2-m_1$.

\subsection{Complex null vectors and roots of unity}

Our calculations     require the use of an other    basis    of ${\cal V} ^{k }$  which was introduced in      \cite{lachs3} (and apparently   ignored before in the literature): \begin{equation} \label{ }
B3\equiv (\Phi^{k}_{IJ}), ~ I,J=0..k. 
\end{equation}    This basis may only be used when $k$ even, which is assumed here. Hereafter, all sums involving $i,j,I$ or  $J$ will cover the range $0,k$.

The basis functions are defined as 
 $\Phi^{k}_{IJ}:~\Phi^{k}_{IJ} (X) \equiv (X \cdot N_{IJ} )^{k}$,  from  the $(k+1)^2$ null vectors of $\setC ^4$,
  \begin{equation}\label{ }
N_{IJ} \equiv N(I \alpha,J\alpha)=(\cos I  \alpha  ,~i~\sin J\alpha ,~i~\cos J \alpha ,~\sin I   \alpha ).\end{equation} 
Here the dot  product extends   the  Euclidean [scalar] dot  product of $\R ^4$ to its  complexification $\setC ^4$. (Complex) \emph{null}  vectors are  defined as having   zero    norm.
  The angle 
 $\alpha  \equiv \frac{2 \pi}{k+1}$
is the argument of   the   $(k +1)^{th}$ complex roots of unity. Those   are the powers $\rho ^I$ of \begin{equation}
\label{}\rho \equiv e^{\frac{2 i \pi }{k+1}} \equiv \cos \alpha +i \sin \alpha  .
\end{equation}   
 
    We proved in \cite{lachs3}
  that B3 form a basis of $V^k$ (when $k$ is even) and we gave the transformation formulae with B2. First we  defined $~Ê{\cal  T}_{k;m_1,m_2} \equiv ~P_{k;m_1,m_2}~ T_{k;m_1,m_2}$, with    $$ P_{k;m_1,m_2}  =\frac{ 2^{-k}~k! ~}{ (k/2- m_1)!~(k/2+  m_1)!~(k+1)^2 ~ C_{k;m_1,m_2}}  $$
   $$ =\frac{ 2^{-k}~~\pi~k! ~(k+1)^{-5/2}}{\sqrt{{ (k/2+m_2)!~(k/2-m_2)!  (k/2+m_1)!~Ê(k/2-m_1)!}}}  .$$
 Then the transformation reads    
      \begin{equation} \label{Tphi}
{\cal  T}_{k;m_1,m_2}= \frac{1}{(k+1)^2 }~\sum _{I,J=0}^k \rho^{I(m_1+m_2)-J (m_2-m_1)}~ \Phi^{k}_{IJ}, \end{equation} 
$$  \Phi^{k}_{IJ} = \sum _{m_1,m_2=-k/2}^{k/2}~ 
{\cal  T}_{k;m_1,m_2}~  \rho^{-I(m_1+m_2)+J (m_2-m_1)}. $$  

\subsection{Rotations in $\R ^4$}

The main interest of the basis B3 lies in the possibility 
 to calculate  explicitly its   rotation properties under an arbitrary rotation $g \in $SO(4). The action of a   rotation $g\in $ SO(4) on a function    is noted   $\mathbf{R}_g$:
$\mathbf{R}_g f(x)\equiv f(gx)$. Applied  to the basis functions, it may be developed as \begin{equation}
\label{ }\mathbf{R}_g:  \Phi_{IJ} \mapsto \mathbf{R}_g \Phi_{IJ}:  \Phi_{IJ} (x)=\Phi_{IJ} (gx)= \sum _{i j=0}^k~ G_{IJ}^{i j}(g)~\Phi_{i j}.\end{equation}  The coefficients were calculated in  \cite{lachs3} as
\begin{equation}
\label{G} G_{IJ}^{ij}= \frac{\left(  {\cal A}'    \right)^{k}}{(k+1)^2} \sum _{A,B=0 }^k~ \rho ^{-i(A+B-k)} ~Ê\rho ^{-j(A-B)}~{\cal U }^{A}
~{\cal V }^{B},\end{equation}
where we defined   ${\cal U }\equiv  \left(\frac{  {\cal B}}{ {\cal A}' }\right)$,
${\cal V} \equiv  \left(\frac{{\cal A} }{    {\cal B}}\right)$,
 ${\cal A} \equiv <Q_L~ \alpha ~Q_R \cdot n_{IJ}>$,
${\cal A}' \equiv <Q_L~ \bar{\alpha} ~Q_R \cdot n_{IJ}>$,
${\cal B} \equiv <Q_L~ \beta ~Q_R \cdot n_{IJ}>$,
${\cal D} \equiv <Q_L~ \delta ~Q_R \cdot n_{IJ}>$ . Note that $A A'=B D$.\\
 In these formulae, the two unit quaternions $Q_L$ and $Q_R$ represent the rotation $g$. The four basis quaternions are written $j_{\mu}$, $j=0,1,2,3$, with $j^0=1$.  A quaternions is developed as $q=q^{\mu}~j_{\mu}$. For a complex quaternions, the $q^{\mu}$ are complex numbers.
   Terms like $Q_L~ \alpha ~Q_R$ simply denote  the quaternionic product. The brackets indicate the quaternionic scalar product, the bar the quaternionic conjugate. 
 The null complex quaternions $n_{IJ}$ represent the null vectors $N_{IJ}$ introduced above (see details in \cite{lachs3}). Also, we introduced  the following peculiar   null complex quaternions\\ $ \alpha \equiv 1+i~j_3$,  $\beta \equiv j_1-i~j_2=  (1-i~j_3)~j_1$ and  $\delta \equiv - j_1-i~j_2$. They have zero norm and obey  the properties\\ $<\alpha \cdot n_{IJ}>=  \rho ^{I},~<\bar{\alpha} \cdot n_{IJ}>=  \rho ^{-I},~<\beta \cdot n_{IJ}>= \rho ^J,~<\delta \cdot n_{IJ}>= \rho ^{-J}$.

    The  coefficients $G_{IJ}^{ij}$  completely define the transformation properties of the basis functions under any element $g$ of SO(4). 

\section{Dodecahedral space}

\subsection{Generators}
 
The  Poincar\'e  Dodecahedral space has two generators $g_{\pm}$, acting as $x \mapsto g_{\pm} x$, for any point of $S^3$ represented by the unit vector $x$ of the embedding space $\R ^4$. In a certain  basis of $\R ^4$,  they   are expressed     (Weeks, prived communication) as the two matrices $g_{\pm} \equiv 
\begin{pmatrix}
     c &-C   &\pm 1/2&0\\
 C  & c&0&\mp 1/2\\
     \mp 1/2 &  0 &c&-C\\ 
 0 & \pm 1/2  &C &   c
     \end{pmatrix}
$,   with $C \equiv \frac{\sqrt{5}-1}{4} $, $c \equiv \frac{\sqrt{5}+1}{4} $. 

It follows    their (left action)  {\sl  complex matrix} forms as
$G_{\pm} \equiv 
\begin{pmatrix}
     c +iC   &\pm 1  /2\\
\pm 1  /2  & c -iC   
     \end{pmatrix}$, whose action is defined by $$ \matrixdd{W}{iZ}{i\bar{Z}}{\bar{W}} \mapsto G_{\pm}  \matrixdd{W}{iZ}{i\bar{Z}}{\bar{W}}; ~W \equiv x^0+ix^3, ~Z \equiv x^1+ix^2 \in \setC.$$

These two operators have the same  eigenvalues, namely  $\lambda  \equiv e^{i\pi/5}$ and $\lambda ^*$ (star means complex conjugation).

The corresponding (unit)  quaternions $Q_{\pm}$  act by left action only (as   $
q_x \mapsto Q_{\pm}~ q_x$, where $q_x$ is the unit quaternion associted to the point $x$ of $S^3$).  They are given by     $Q_{\pm}= c+C~j_3 \mp ~j_1/2$. We note $\dot{\lambda } \equiv e^{j_3 \pi/5}$ the quaternionic analog of the eigenvalue $\lambda$.

{\bf Diagonalisation}

Calculations are   much   easier if we adopt  a different  basis  where one  generator, say $g_+$, 
transform the complex coordinates $W \equiv x^0+ix^3$ and $Z \equiv x^1 + ix^2$, of an arbitrary vector  $x \in \R^3$,  by scalar (although complex) multiplication. To find such a basis  is equivalent to  diagonalize the complex matrix $G_+$, what we have done. We express the results in quaternionic notation. We note $u$ the quaternion which expresses the change of coordinates: by definition, $\dot{\lambda } =u~Ê Q_+  ~Êu ^{-1}$. Calculations give
$$u=1/2+j_2~(\sqrt{C^2+1/4}-C).$$ The second generator takes the form  \begin{equation}
\label{second}
ÊR_-  \equiv u~Ê Q_- ~Ê u ^{-1}=\cos(\pi /5)Ê-j_3~\sin (\pi /5)~/\sqrt{5} + Êj_1~2 ~Ê\sin (\pi /5)~/\sqrt{5}.
\end{equation}
  Now we   continue the calculations with this  basis:  the two    generators are $  \dot{\lambda } $ (diagonal)  and $R_-$  (formula \ref{second}).

\subsection{Invariance}

The diagonal character  of the  first generator makes the calculations  easy: the basis functions $T_{k;M_1, M_2}$ of B2 appear to be  its eigenfunctions. This is analog to    the case of a  lens space, as seen in \cite{lachs3}. Thus, the functions of $V^k$ which are   $g_+$ invariant are all  combinations of  the $T_{k;m_1,m_2}$ which verify the condition $2 m_2=0 \mod{10}$. This implies 
$k$ even  (as was derived by \cite{Ikeda}), so  that $m_1 $ and $m_2 $ are entire and \begin{equation}
\label{condition}\underline{m_2}=0 \mod 5,
\end{equation} where we  underline as $\underline{m_2}$ a value of $m_2$ verifying this condition. It remains to express the invariance condition with respect to the second generator.
 
{\bf Second generator}

We assume   $k$  even, and we adopt  the basis B3 of $V^k$. Simple calculations lead to estimate \begin{equation} \label{ }
{\cal A}_{IJ} \equiv <(R_-) \alpha ~  \cdot n_{IJ}>=F~\rho ^I + E~\rho ^J,~{\cal A}'_{IJ} \equiv <(R_-) \bar{\alpha} ~  \cdot n_{IJ}>=F^*~\rho ^{-I}- E~\rho ^{-}J,
\nonumber \end{equation}
 \begin{equation}
\label{AB}
{\cal B}_{IJ} \equiv <(R_- )^P\beta ~ \cdot n_{IJ}>=F^*~\rho ^J -E_P~\rho ^I,
\end{equation}
with $F \equiv \cos(\pi /5)+i~ \sin(\pi /5)/\sqrt{5},$ 
$E \equiv 2 ~\sin(\pi /5)/\sqrt{5}$.

We insert the relation
 ${\cal U }\equiv  \frac{  {\cal B}}{ {\cal A}' }=\rho ^{I+J}$  in (\ref{G}). Taking into account the properties of the roots of unity,  we obtain
$$G_{IJ}^{ij}(R_-)~= \frac{ ({\cal A}'   )  ^{k}~ \rho ^{-i} }{ (k+1)}~\delta ^{Dirac}~\frac{{\cal V }^{k+1}-1}{{\cal V }~\rho ^{I+J-2i}-~1},$$ 
where $\delta ^{Dirac}$ holds for $\delta ^{Dirac}[(I+J)-(i+j),~\mod(k+1)]$.

This expression allows to return to  the rotation properties of the  basis B2 (in fact ${\cal T}_{k;M_1, M_2}$ rather than $ T _{k;M_1, M_2}$), that we express by the development  \begin{equation}
\label{developT}
\mathbf{ R}_{g_-}  {\cal T}_{k;M_1, M_2} \equiv \sum _{m_1 ,m_2}~ \Gamma_{M_1 M_2}^{m_1 m_2}(g)~{\cal T}_{k;m_1, m_2}.\end{equation}
The coefficients 
take the form\begin{equation}\label{ }
\Gamma_{M_1 M_2}^{m_1 m_2}=\sum_{i j}~\sum_{IJ}~Ê\rho ^{I (M_1+M_2)+J (M_1-M_2) -i(m_1+m_2)-j~ (m_1-m_2)}~\frac{G_{IJ}^{ij}}{(k+1)^2}~
\end{equation}
$$= \sum_{IJ}~Ê\rho ^{(I -J)~M_2+ (I+J)~ (-m_1+m_2+M_1)}
\frac{({\cal V }^{k+1} -1)~({\cal A}')^k}{~(k+1)^3}~\sum_{i }~\frac{\rho^{-2i~ m_2}   }{ ~({\cal V }~\rho ^{I+J-i}-~\rho ^{i})}.$$
Direct calculations allow to evaluate the last sum as $$ (k+1)~\frac{({\cal V }~ÊÊ\rho ^{I +J} )^{k/2-m_2}}{{\cal V }^{k+1}-1 },$$ leading finally to
\begin{equation} \label{ }\Gamma_{M_1 M_2}^{m_1 m_2}= 
 \sum_{IJ}~Ê\rho ^{(I-J)~ M_2 +(I+J)~ (M_1-m_1+k/2)}~
\frac{{\cal V }^{k/2-m_2} ~({\cal A}')^k}{~(k+1)^2}
.\end{equation}

{\bf Still a new basis}

This suggests   us the  introduction of still a new basis for $V^k$ (only for intermediary calculations): \begin{equation}
\label{ }
\tilde{T}_{\alpha M_2} \equiv \sum _{M_1}~\rho ^{-\alpha ~M_1}~{\cal T}_{k;M_1, M_2},~\alpha =0..k,~ M_2=-k/2..k/2,\end{equation} with inverse  formula $ {\cal T}_{k;M_1, M_2}= \frac{1}{k+1}  \sum _{\alpha}~\rho ^{\alpha ~M_1}~\tilde{T}_{\alpha M_2}$.  Its rotation properties  are  deduced as:\begin{equation} \label{ }
\mathbf{ R}_{g_-}: \tilde{T}_{\alpha M_2}  \mapsto \frac{\rho ^{\alpha ~Ê(k/2-M_2)}}{k+1}  ~\sum _I~Ê 
 \rho ^{2I~M_2}~({\cal A}')^k ~{\cal V }^{k/2} \sum _{m_2} {\cal V }^{-m_2}
~Ê\tilde{T}_{\alpha m_2}, \end{equation}
where the quantities ${\cal A}'$ and  ${\cal V }$ have to be evaluated for the value $J=\alpha -I \mod (k+1)$. 

A first     important result  appears:   the value of $\alpha$, for this basis, is  preserved  by $\mathbf{ R}_{g_-}$. Using the formula just above, it is straightforward to check that this implies, similarly, that  $\mathbf{ R}_{g_-}$ preserves  $m_1$   in the basis B2. This result will  allow  considerable simplification. To continue, we   report in the  formula above the values of    $   {\cal V }={\cal A }/{\cal B }$ and ${\cal A}'$ (equ.\ref{AB}). This gives  $$\mathbf{ R}_{g_-}: \tilde{T}_{\alpha M_2}  \mapsto   \sum _{m_2}~\gamma _{\alpha M_2}^{m_2}~Ê
~Ê\tilde{T}_{\alpha m_2},$$ with 
 \begin{equation}\label{ }\gamma _{\alpha M_2}^{m_2}=\sum _I~Ê 
\frac{ \rho ^{(I-\alpha/2)~(2M_2+k)} }{k+1}~ 
 [F+E~\rho ^{ \alpha -2I}]^{k/2-m_2}
[F*~Ê\rho ^{\alpha -2I}-E]^{k/2+m_2},\end{equation}
where    $k/2-m_2$ and $k/2+m_2$  take their value between  0 and $k$. 

\subsection{Splitting the eigenspace}

From  this formula, it appears   the second important result   that
 \begin{equation}\label{gamma}\gamma _{\alpha M_2}^{m_2}=\gamma _{M_2}^{m_2}=\sum _I~Ê 
\frac{ \rho ^{I~(2M_2+k)} }{k+1}~    [F+E~\rho ^{  -2I}]^{k/2-m_2}
[F*~Ê\rho ^{ -2I}-E]^{k/2+m_2}\end{equation} 
does not depend on  $\alpha$. This allows to write the rotation formula
$$\mathbf{ R}_{g_-}: \tilde{T}_{\alpha M_2}  \mapsto   \sum _{m_2}~\gamma _{ M_2}^{m_2}~Ê
~Ê\tilde{T}_{\alpha m_2}.$$   
It is easily checked that this  absence of dependence on  $\alpha$ holds also for the ${\cal T}_{k;m_1, m_2}$. Finally, we are led  to    the  transformation rule for the basis B2: \begin{equation}
\label{mmm}\mathbf{ R}_{g_-} : {\cal T}_{k;M_1,M_2} \mapsto \sum _{m_2} \gamma _{M_2}^{m_2}~Ê{\cal T}_{k;M_1,m_2},\end{equation}
with $ \gamma _{M_2}^{m_2}$ given by (\ref{gamma}).

{\bf Summary}

We summarize the  results obtained,  expressed in the basis $({\calT}_{k;m_1,m_2})$:\begin{itemize}
  \item 
  The    $g_+$ invariant functions of $V^k$  are   all  combinations of  the ${\calT}_{k;m_1,\underline{m_2}}$ which verify the condition
  $\underline{m_2}=0 \mod 5$.  
  \item 
The second generator  ($g_-$)  preserves the value of $m_1$. This allows to   consider $V^k$ as the direct sum of $k+1$ sub vector-spaces $V^{k,m_1}$ of dimension $k+1$, each  preserved by $g_-$. The search for invariant functions can thus be made independently for each  $V^{k,m_1}$.\\   \item   
The rotation coefficients (\ref{gamma}) in $V^{k,m_1}$  do not depend on $m_1$: they  are identical in  each of the  $V^{k,m_1}$. 
It is thus sufficient to search   invariant functions in one of them (say,   for $m_1 =0$).  The other functions are  given by the  $k$ more copies obtained by replacing the value $m_1 =0$ by any of the $k$ remaining values. It results that  the dimension of the space of eigen functions of the dodecahedral space is  an entire multiple of $k+1$, as was already indicated by \cite{Ikeda}.
\end{itemize}

\subsection{  Eigenvectors}

The search of the modes of the dodecahedral space  is reduced (for each even value of $k$) to that  of the invariant eigenfunctions of $S^3$  in $V^{k,m_1=0}$. Invariance with respect to  $g_+$ implies that   such a function may decomposed in the restricted basis formed by  the ${\cal T}_{M_1=0 ,{\underline M_2}}$ (we recall that underlining means ${\underline M_2}=0 \mod 5$). In this restricted basis, the rotations properties  (under $g_-$)   are expressed by the $k_5 * k_5$ matrix $\mathbf{ G}_k$ of coefficients  $\gamma _{\underline M_2}^{\underline m_2}$ given by  (\ref{gamma})
(we define $k_5 \equiv 1+2 [k/10]$). Thus,  invariant functions are  the eigenvector(s) of the matrix  $\mathbf{ G}_k$,   corresponding to the eigenvalue 1, when they exist.  We  note such an eigenvector by its  $k_5$ components 
$(f _{\underline{M_2}})$. (Note that, when  the multiplicity of the eigenvalue 1 is larger than 1, they can be several such vectors; this occurs for instance for the value $k=60$, see the table.)  
Finally, \\

{\framebox{\shortstack{
the eigenfunctions of the dodecahedral space corresponding to the \\ eigenvalue $-k~(k+2)$ ($k$ even) are all combinations of the functions \\
$\sum _{\underline{M_2}}~Êf _{\underline{M_2}}~Ê{\cal T} _{k;M_1,\underline{M_2}},$\\
where $(Êf _{\underline{M_2}})$ is an eigenvector of the matrix $\mathbf{ G}_k$ with eigenvalue 1, \\ for all the entire values of    $M_1 $ between 0 and $k$ .
}}}

For a value of $k$, the calculation of the   $(k_5)^2$ coefficients of $\mathbf{ G}_k$  is immediate. The search for eigenvectors also runs very easily  (on MAPLE for instance). Results  confirms the values of the eigenvalues  and their multiplicities given by  \cite{Ikeda}.   The table   gives the numerical values of the first  eigenvectors  of the dodecahedron.  MAPLE codes to calculate the modes may be obtained by request at marclr@cea.fr.

\npage {\bf Table 1}: For each value of $k$ (from 12 to 44;  eigenmodes exist only for the  values given), the table give the $k_5$ components ($k_5 \equiv 1+2 [k/10]$) of the vector $f _{\underline{M_2}}$. The 
eigenmodes of D, corresponding to $\lambda _k=-k~(k+2)$, are given by all combinations
$\sum _{\underline{M_2}}~Êf _{\underline{M_2}}~Ê{\cal T} _{k;M_1,\underline{M_2}},$
where $\underline{M_2}$ varies  from  $-k/2$ and $k/2$ and  verifies (\ref{condition}), and $M_1$ takes all the entire values between $-k/2$ and $k/2$.\\

\small{
{\bf   k=12}:~Ê
.70352647068144845281,~
-.10050378152592120757~i,
~ .70352647068144845294\\

{\bf   k= 20}:~
.70702906968084266661,~
.010397486318835921579~i,\\
~ -.0018904520579701675214, 
.010397486318835921658~i,
~ .70702906968084267112\\

{\bf   k=24}:~
.70697192267503596680,
~-.012403016187281332779~i,\\
 -.85867035142716919288~e-2,
~-.012403016187281332763~i,
~ .70697192267503596704\\

{\bf   k= 30}:~
~.70709634963126361072+.28263381117495907229~e-2~i,\\
 ~.10352886856225736179~e-4-.25900965187958376464~e-2~i,\\
 ~.23546331989053068040~e-3+.94117153238415742893~e-6~i,
~0,\\
 ~-.23546331989053075280~e-3-.94117153238413699545~e-6~i,\\
 ~-.10352886856225582837~e-4+.25900965187958382323~e-2~i,\\
 ~-.70709634963126377938-.28263381117495713938~e-2~i\\

{\bf   k= 32}:~
.70708591611133287319, 
~.54182828820791792693~e-2~i, \\
 ~.35336627491820729542~e-3,
~.22172001563495361997~e-3~i,\\
 ~.35336627491820731478~e-3,
~.54182828820791791292~e-2~i,\\
~ .70708591611133284832\\

{\bf   k=36}:~
.70708473425088583343, -.55056653389946616956~e-2~i,\\
~ -.78652361985638027430~e-3,~Ê
~.70373165987149811020~e-3~i,\\
~ -.78652361985638024036~e-3,
~-.55056653389946617208~e-2~i,\\
 ~.70708473425088584255\\

{\bf   k= 40}:~
.70710661000066274387 ,
~.49002537075583660713~e-3~i,\\
 ~-.44188504840885072726~e-4,
~-.39518175060953038014~e-5~i,\\
~ .71851227383553307597~e-6,
-.39518175060954735563~e-5~i,\\
~ -.44188504840885377434~e-4,
~.49002537075583232202~e-3~i,\\
 ~.70710661000067363741\\

{\bf   k= 42}:~
.70710034184850288923,
~-.30174979026251895955~e-2~i,\\ ~ .29583312770835158828e-4
~-.19722208513890132964~e-4~i,
~0\\
~.19722208513890135205~e-4~i,
 ~-.29583312770835191330~e-4\\
~.30174979026251928239~e-2~i,
 -.70710034184850303513\\

{\bf   k= 44}:~
.70709762382971759036,
~.35958850229318122280~e-2~i,\\
 ~.13964602030803151449~e-3,
~-.15649985034520771577~e-4~i,\\
~ .21878574396085900270~e-4,
~-.15649985034520774732~e-4~i,\\
~ .13964602030803151858~e-3,
~.35958850229318113024~e-2~i,\\
 ~.70709762382971762555 \\
}\npage {\bf Table 1, continued}: idem, for   values of $k$ from  48 to 56\\

\small{
{\bf   k= 48}:~
.70709753381495795208
~-.36096867863497570254~e-2~i,\\
~ -.20595810492686306199~e-3,
~.61892333738632857047~e-4~i,\\
 ~.56297546508021977011~e-4
~.61892333738632854996~e-4~i,\\
 ~-.20595810492686303733~e-3,
~-.36096867863497572060~e-2~i,\\
~ .70709753381495797378\\

{\bf   k=50}:~
.70710677432308054872 ,
~-.98118423819072486562~e-4~i,\\
 ~.88473375505798110701~e-5
~.79771076275668425524~e-6~i,\\
~ -.72519160250664405750~e-7
~.13686767861056995799~e-19~i,\\
~ .72519160250661739763~e-7,
~-.79771076275733196796~e-6~i,\\
 ~-.88473375505808361189~e-5,
~.98118423819152347705~e-4~i,\\
 ~-.70710677432345667134 \\

{\bf   k=52}:~
.70710632382180878363,
~.80371462823607746218e-3~i,\\
 ~-.29191620936716760517e-4,
~.12714219501894409823e-5~i,\\
~ -.45691726334935131233e-6,
~-.14757576207556568245e-6~i,\\
~ -.45691726334935055995e-6,
~.12714219501895343589e-5~i,\\
~ -.29191620936716944313e-4,
~.80371462823612059084e-3~i,\\
 ~.70710632382183246593 \\

{\bf   k= 54}:~
.70709995013538603217,
~-.31081316489467430301~e-2~i,\\
 ~-.47140520459254381578~e-5,
~-.33908093663673232699~e-5~i,\\
~ -.15713506819751032060~e-5,
~0,\\
 ~.15713506819751014572~e-5 ,
~.33908093663673239606~e-5~i,\\
 ~.47140520459254356244~e-5 ,
~.31081316489467431476~e-2~i,\\
 ~-.70709995013538374766 \\
 
{\bf    k=56}:~
 .70710165828128236325,
~.26907309019741289751~e-2~i,\\
~ .69176592337987433371~e-4,
~-.59344090988012596170~e-5~i,\\
~ -.63998529496876216108~e-6,
~-.20222732831614883787~e-5~i,\\
~ -.63998529496876549920~e-6,
~-.59344090988012761721~e-5~i,\\
~ .69176592337987085779~e-4,
 ~.26907309019741259726~e-2~i,\\
 ~.70710165828128410534\\
 }\npage {\bf Table 1, continued}: idem, for   values of $k$ from  60 to 62\\

\small{
{\bf   k= 60}:~
 6.9394960301822869967-.50386496302764001912~i,\\
~ .29358129642425900098-.64161677487791106593~i,\\
 ~-.25742682420776540173e-2-.11683763224239293660~e-2~i,\\
 ~-.36432453284660713864e-4+.77853455569667120805~e-4~i,\\
 ~.11085720861501942161e-4+.49965917891592449037~e-5~i,\\
 ~.21499250718118950501e-5-.46857540636509975228~e-5~i, \\
 ~-.39964490355299009740e-5-.18268003343521902658~e-5~i,\\
 ~ .21499250718136508514e-5-.46857540636526609287~e-5~i,\\
 ~.11085720861498572522e-4+.49965917891578880069~e-5~i,\\
 ~-.36432453284628003387e-4+.77853455569638907558~e-4~i,\\
 ~-.25742682420783301674e-2-.11683763224246853776~e-2~i,\\
 ~.29358129642426822804-.64161677487790595997~i,\\
~ 6.9394960301835349645-.50386496298031253450~i\\

.80848705655829822065e-2+.70706052583861537173~i,\\
 -.20832969516412067927e-3+.65282867307451706578~e-4~i,\\
~ .23918191153754173532e-6-.99923368048573892481~e-6~i,\\
~ .18133484953835426669e-6-.98554877105180951336~e-8~i,\\
~ -.94672012551906538764e-9+.11022887144687929246~e-7~i,\\
~ -.26567291467002942075e-8+.49083380949540372946~e-9~i,\\
~ .40224546881829871331e-9+.94397550786428917031~e-9~i,\\
~ -.26567291466915270685e-8+.49083380949346250026~e-9~i,\\
~ -.94672012552856441064e-9+.11022887144643159845~e-7~i,\\
~ .18133484954122299052e-6-.98554877083976930526~e-8~i,\\
~ .23918191155535933834e-6-.99923368050776437049~e-6~i,\\
~ -.20832969516570937280e-3+.65282867306944956583~e-4~i,\\
~ .80848705598891417065e-2+.70706052584512272298~i\\

{\bf   k=  62}:~
.70710674764980394726,
~-.21751107313340935062~e-3~i,\\
~ .10807153948145142931~e-4,
~.18045880549024964468~e-6~i,\\
 ~.55301892005028276440~e-7,
~.11642503580014562160~e-7~i,
~0,\\
~-.11642503580010046955~e-7~i,
~ -.55301892005094596638~e-7,\\
~-.18045880548966284252~e-6~i,
 ~-.10807153948148682328~e-4,\\
~.21751107313369991548~e-3~i,
~ -.70710674765011395485]

\end{document}